\documentclass[aps,prd,preprint,floatfix,nofootinbib,superscriptaddress]{revtex4}
\usepackage{graphicx,subfigure,epsfig,epsf}

\def\poutsla{/\!\!\!p_{out}}
\def\pinsla{/\!\!\!p_{in}}

\newcommand{\beq}{\begin{equation}}
\newcommand{\eeq}{\end{equation}}
\newcommand{\bea}{\begin{eqnarray}}
\newcommand{\eea}{\end{eqnarray}}

\newcommand{\Lda}{\Lambda}
\newcommand{\g}{\gamma}
\def\Re{{\cal R \mskip-4mu \lower.1ex \hbox{\it e}\,}}
\def\Im{{\cal I \mskip-5mu \lower.1ex \hbox{\it m}\,}}

\def\etal{{\it et al.}}

\def\tev{\,{\ifmmode\mathrm {TeV}\else TeV\fi}}
\def\gev{\,{\ifmmode\mathrm {GeV}\else GeV\fi}}
\def\mev{\,{\ifmmode\mathrm {MeV}\else MeV\fi}}
\def\to{\rightarrow}

\begin{document}


\def\issue(#1,#2,#3){#1 (#3) #2} 
\def\APP(#1,#2,#3){Acta Phys.\ Polon.\ \issue(#1,#2,#3)}
\def\ARNPS(#1,#2,#3){Ann.\ Rev.\ Nucl.\ Part.\ Sci.\ \issue(#1,#2,#3)}
\def\CPC(#1,#2,#3){Comp.\ Phys.\ Comm.\ \issue(#1,#2,#3)}
\def\CIP(#1,#2,#3){Comput.\ Phys.\ \issue(#1,#2,#3)}
\def\EPJC(#1,#2,#3){Eur.\ Phys.\ J.\ C\ \issue(#1,#2,#3)}
\def\EPJD(#1,#2,#3){Eur.\ Phys.\ J. Direct\ C\ \issue(#1,#2,#3)}
\def\IEEETNS(#1,#2,#3){IEEE Trans.\ Nucl.\ Sci.\ \issue(#1,#2,#3)}
\def\IJMP(#1,#2,#3){Int.\ J.\ Mod.\ Phys. \issue(#1,#2,#3)}
\def\JHEP(#1,#2,#3){J.\ High Energy Physics \issue(#1,#2,#3)}
\def\JPG(#1,#2,#3){J.\ Phys.\ G \issue(#1,#2,#3)}
\def\MPL(#1,#2,#3){Mod.\ Phys.\ Lett.\ \issue(#1,#2,#3)}
\def\NP(#1,#2,#3){Nucl.\ Phys.\ \issue(#1,#2,#3)}
\def\NIM(#1,#2,#3){Nucl.\ Instrum.\ Meth.\ \issue(#1,#2,#3)}
\def\PL(#1,#2,#3){Phys.\ Lett.\ \issue(#1,#2,#3)}
\def\PRD(#1,#2,#3){Phys.\ Rev.\ D \issue(#1,#2,#3)}
\def\PRL(#1,#2,#3){Phys.\ Rev.\ Lett.\ \issue(#1,#2,#3)}
\def\PTP(#1,#2,#3){Progs.\ Theo.\ Phys. \ \issue(#1,#2,#3)}
\def\RMP(#1,#2,#3){Rev.\ Mod.\ Phys.\ \issue(#1,#2,#3)}
\def\SJNP(#1,#2,#3){Sov.\ J. Nucl.\ Phys.\ \issue(#1,#2,#3)}


\bibliographystyle{revtex}

\title{Effect of earth rotation on pair production of Standard Model Higgs bosons at linear colliders in  the noncommutative space-time } 



\author{Prasanta~Kumar~Das}
\email[]{Author(corresponding) : pdas@bits-goa.ac.in}
\affiliation{Birla Institute of Technology and Science-Pilani, K. K. Birla Goa campus, NH-17B, Zuarinagar, Goa-403726, India }
\author{Abhishodh Prakash}
\email[]{Abhishodh.Prakash@stonybrook.edu}
\affiliation{Department of Physics and Astronomy, Stony Brook University, Stony Brook, New York 11790, USA}


\date{\today}

\begin{abstract} 
{We study the neutral Higgs boson pair production through $e^+~e^-$ collision in the noncommutative(NC) extension of the standard model using the Seiberg-Witten maps of this to the first order of the noncommutative parameter $\Theta_{\mu \nu}$. This process is forbidden in the standard model at the tree level with background space-time being commutative. After including the effects of earth's rotation we analyse the time-averaged cross section of the pair production of Higgs boson (in the light of LEP II and LHC data) at the future Linear Collider which can be quite significant for the NC scale $\Lambda$ lying in the range $0.3 - 1.0$ TeV. For the $125$ GeV Higgs mass(the most promising value of Higgs mass as reported by LHC), we find the NC scale as $330~ \rm{GeV}$ $\le \Lambda \le 660 ~\rm{GeV}$ and using $m_h = 129(127.5) ~\rm{GeV}$ (the lower threshold value of the excluded region of $m_h$ reported by ATLAS(CMS) collaborations of LHC), we find the bound on $\Lambda$ as: (i)~$339~(336) ~\rm{GeV} \le \Lambda \le 677~(670) ~\rm{GeV}$ corresponding to the Linear Collider energy $E_{com} = 500 ~\rm{GeV}$.} \\
{\bf PACS No:} 11.10.Nx, 12.60.-i, 14.80.Ec\\
{\bf Keywords:} Noncommutative spacetime, earth rotation, Higgs boson, scattering cross section.
\end{abstract}

\maketitle


\section{Introduction }
The complete experimental success of the Standard Model(SM) of Particle Physics strongly resides on the discovery  of the unseen particle, the Higgs boson\cite{Higgs}. After the Large Electron Positron (LEP) collider has set a lower limit of about 114.3 GeV on its mass \cite{TJ}, the responsibility of finding the Higgs fell on the Large Hadron Collider (LHC) at CERN. Recently two LHC collaborations, ATLAS and CMS from their initial measurement excluded the Standard Model(SM) Higgs mass at the $95~\%$ confidence level from $129~(127.5)~\rm{GeV}$ to $539~(600)~\rm{GeV}$\cite{ATLASCMS}. Both ATLAS and CMS has found strong hints of SM like Higgs boson at a mass $125~\rm{GeV}$ with local significances of $2.5~\rm{\sigma}$ and $2.8~\rm{\sigma}$ which has been updated to $4.9~\sigma$ on $4~\rm{th}$ July,2012 \cite{Joe}.
So, wait few more hours, may be 2-3 days to hear the last set of words of Higgs discovery from ATLAS and CMS  collaborations at CERN. 

 Although the discovery of Higgs boson is the most important activity of the LHC at CERN and the particle  phenomenologists around the world, at the same time there are several issues associated with the Higgs boson that need to be resolved. One such problem is the naturalness problem which make a strong case for physics beyond the standard model(SM), just around or above the mass scale where the Higgs boson is expected to be found. It is therefore of supreme interest to see if the collider signals of the Higgs boson contain some imprint of new physics. This necessitates detailed quantitaive exploration of a variety of phenomenona linked to the production and decays of the Higgs. 

In this paper, we have studied pair production of the SM like Higgs boson in the light mass range ($114.3 ~\rm{GeV} - 129~\rm{GeV}$) at the Linear Collider(LC) as a possible channel for uncovering new physics effects.  In particular, we show that the pair production of Higgs boson which is forbidden in the SM at the tree level receives a large contribution in the noncommutative(NC) extension of the SM. 

As mentioned above, the two widely separated scales: the electroweak scale $M_{W}$ and the Planck scale $M_{\it Pl}$ leads to the notorius Higgs hieracrhy and fine tunning problem which are quite puzzling. Though theories like supersymmetry and technicolour, with their own phenomenological implications and constraints, have been proposed to resolve this problem, the idea of extra spatial dimensions with the scale of gravity being as low as TeV, has drawn a lot of interest among the physics community \cite{VR}. In a class of brane-world models \cite{ADD98} where this TeV scale gravity is realised, one can principly expect to see some stringy effects and the signature of space-time noncommutavity in the upcoming TeV energy colliders like Large Hadron Collider, Linear Collider etc. 

 A lot of interests in the noncommutative(NC) field theories arose from the pioneering work by Snyder \cite{Snyder47}. It has drawn further attention recently due to developments connected 
to string theories in which the noncommutativity of space-time is an important characteristic of D-Brane dynamics at low energy limit\cite{Connes98,Douglas98,SW99}. Although Douglas \etal \cite{Douglas98} in their pioneering work have shown that noncommutative field theory is a well-defined quantum field theory, the question that remains is whether the string theory prediction and the noncommutative effect can be seen at the energy scale attainable in present or near future experiments instead of the $4$-$d$ Planck scale $\rm{M_{pl}} \sim 10^{19}~\rm{GeV}$. A notable work by Witten \etal \cite{Witten96} suggests that one can see some stringy effects by lowering down the threshold value of commutativity to \tev, a scale which can be probed at the present running LHC and at the upcoming collider like Linear Collider(LC).  

~ What is meant by the noncommutative space-time?  It means space and time no longer commute with each other and so one cannot measure the space and time coordinates simultaneously with the same accuracy. Writing the space-time coordinates as operators  we find 
\beq 
[\hat{X}_\mu,\hat{X}_\nu]=i\Theta_{\mu\nu}
\label{NCSTh}
\eeq
where the matrix $\Theta_{\mu\nu}$ is real and antisymmetric. The NC parameter $\Theta_{\mu\nu}$ has dimension of area and reflects the extent to which the space-time coordinates are noncommutative i.e. fuzzy. Furthermore, introducing a NC scale  $\Lda$, we rewrite Eq. \ref{NCSTh} as 
\beq 
[\hat{X}_\mu,\hat{X}_\nu]=\frac{i}{\Lda^2} c_{\mu\nu}
\label{NCST}
\eeq
 where $\Theta_{\mu\nu}(=c_{\mu \nu}/\Lda)$ and $c_{\mu\nu}$ has the same properties as $\Theta_{\mu\nu}$. To study an ordinary field theory in such a noncommutative fuzzy space, one replaces all ordinary products among the field variables with Moyal-Weyl(MW) 
\cite{Douglas} $\star$ products defined by
\begin{equation}
(f\star
g)(x)=exp\left(\frac{1}{2}\Theta_{\mu\nu}\partial_{x^\mu}\partial_{y^\nu}\right)f(x)g(y)|_{y=x}.
\label{StarP}
\end{equation}
Using this we can get the NCQED Lagrangian as
\begin{equation} \label{ncQED}
{\cal L}=\frac{1}{2}i(\bar{\psi}\star \gamma^\mu D_\mu\psi
-(D_\mu\bar{\psi})\star \gamma^\mu \psi)- m\bar{\psi}\star
\psi-\frac{1}{4}F_{\mu\nu}\star F^{\mu\nu} \label{NCL},
\end{equation}
which are invariant under the following transformations 
\bea
\psi(x,\Theta) \to \psi'(x,\Theta) &=& U \star \psi(x,\Theta), \\
A_{\mu}(x,\Theta) \to A_{\mu}'(x,\Theta) &=& U \star A_{\mu}(x,\Theta) \star U^{-1} + \frac{i}{e} U \star \partial_\mu U^{-1},
\eea
where $U = (e^{i \Lambda})_\star$. In the NCQED lagrangian (Eq.\ref{ncQED})
$D_\mu\psi=\partial_\mu\psi-ieA_\mu\star\psi$,$~~(D_\mu\bar{\psi})=\partial_\mu\bar{\psi}+ie\bar{\psi}\star
A_\mu$, $~~ F_{\mu\nu}=\partial_{\mu} A_{\nu}-\partial_{\nu}
A_{\mu}-ie(A_{\mu}\star A_{\nu}-A_{\nu}\star A_{\mu})$. 

 An alternative approach is the Seiberg-Witten(SW)\cite{SW99,Douglas98,Connes98,Jurco} 
approach in which both the gauge parameter $\Lambda$ and the gauge field $A^\mu$
is expanded as 
\bea \label{swgp}
\Lambda_\alpha (x,\Theta) &=& \alpha(x) + \Theta^{\mu\nu} \Lambda^{(1)}_{\mu\nu}(x;\alpha) + \Theta^{\mu\nu} \Theta^{\eta\sigma} \Lambda^{(2)}_{\mu\nu\eta\sigma}(x;\alpha) + \cdot \cdot \cdot 
\eea
\bea \label{swgf}
A_\rho (x,\Theta) &=& A_\rho(x) + \Theta^{\mu\nu} A^{(1)}_{\mu\nu\rho}(x) + \Theta^{\mu\nu} \Theta^{\eta\sigma} A^{(2)}_{\mu\nu\eta\sigma\rho}(x) + \cdot \cdot \cdot
\eea
When the field theory is expanded in terms of the above power series one ends up with an infinite tower of higher dimensional operators which renders the theory nonrenormalizable. However, the advantage is that this construction can be applied to any gauge theory with arbitrary matter representation. In the other approach, called the Weyl-Moyal apprach, the group closure property is only found to hold for the $U(N)$ gauge theories and the matter content is found to be in the (anti)-fundamental and adjoint representations. 
Using the SW-map, Calmet \etal \cite{Calmet} first constructed a model with noncommutative gauge invariance which was close to the usual commuting  Standard Model(CSM) and is known as the {\it minimal} NCSM(mNCSM) in which they listed several Feynman rules comprising NC interaction. Intense phenomenological searches \cite{Hewett01} have been made to unravel several interesting features of this mNCSM. Hewett \etal explored several processes e.g. 
$e^+ e^- \to e^+ e^-$ (Bhabha), $e^- e^- \to e^- e^-$ (M\"{o}ller), 
$e^- \g \to e^- \g$, $e^+ e^- \to \g \g$ (pair annihilation), $\g \g \to e^+ e^-$ and $\g \g \to \g \g$ in context of NCQED. 
Recently, in a work \cite{pdas} we have investigated the impact of $Z$ and photon exchange in the Bhabha and the M\"{o}ller scattering and the impact of space-time noncommutativity(with and withour the effect of earth rotation) in the the muon pair production at LC 
\cite{abhishodh}. Now in a generic NCQED the triple photon vertex arises to order ${\mathcal{O}}(\Theta)$, which however is absent in this minimal mNCSM. Another formulation of the NCSM came in forefront through the pioneering work by  Melic \etal \cite{Melic:2005ep}
where such a triple neutral gauge boson coupling \cite{Trampetic} appears naturally in the gauge sector. We will call this the non-minimal version of NCSM or simply NCSM. In the present work we will confine ourselves within this non-minimal version of the NCSM and use the Feynman rules for interactions given in Melic \etal \cite{Melic:2005ep}.

 In Sec. II we present the cross section of $e^+ e^- \to \gamma, Z \to H H $ in the NCSM, a process which is forbidden in the SM at the tree level. In Sec. III, we make detailed numerical analysis of the pair production cross section, angular distribution. We discuss the prospects of TeV scale noncommutative geometry in this section. Finally, we conclude in Sec. IV.  
\section{Higgs pair production at the future linear collider }
\begin{figure}[htbp]
\vspace{5pt}
\centerline{\hspace{-3.3mm}
{\epsfxsize=15cm\epsfbox{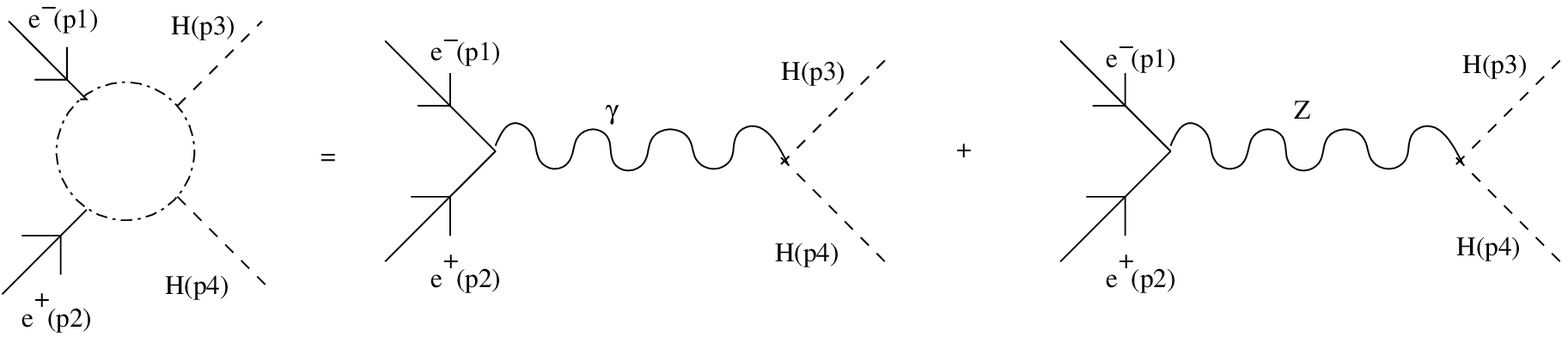}}}
\caption{Feynman diagrams for $ e^- e^+ ~\stackrel{\gamma,Z}{\longrightarrow} ~ H~H $ in the NCSM are shown.}
\protect\label{feyn}
\end{figure}

Since there is no direct coupling of a photon($\gamma$) or a $Z$ boson with a pair of Higgs boson, the pair production of Higgs boson through $e^+ - e^-$ annihilation is forbidden in the SM at the tree level. So an excess in the predicted event rate may be interpreted as the signature of new physics. Supersymmetry and the extra dimensional models are front runners ( see \cite{pkdas} and references therein). 
Here we explore the potential feasibility of the nmNCSM scenario where the pair production proceeds via $s$ channel processes $e^- e^+ \to  \gamma, Z \to H ~ H $. In one of our earlier work \cite{pkdas_higgs}, we have investigated in detail the impact of space-time noncommutativity on the Higgs pair production without considering the effect of earth rotation. Here we include the effect of earth's rotation. The Feynman diagrams for the above process are shown in Fig. \ref{feyn}. 

 The scattering amplitude for the above process using $\mathcal{O} (\Theta)$ Feynman rules (See Appendix A for details ) can be written as 
\bea
i \mathcal{A} = i \mathcal{A}_{\gamma} + i \mathcal{A}_{Z}
\eea
where $\mathcal{A}_{\gamma}$ and $\mathcal{A}_{Z}$ can be calculated as 
\bea \label{unpol_photon}
{i\mathcal{A}_\gamma} = \frac{\pi \alpha ~ m_h^2}{s}  \left[{\overline v}(p_2) \gamma_\mu  u(p_1)\right]  \times (k \Theta)^\mu \left[1 + \frac{i}{2} (p_2 \Theta p_1)\right], 
\eea
\bea \label{unpol_Z}
{i\mathcal{A}_Z} = \frac{\pi \alpha ~ m_h^2}{\sin^2(2\theta_W) s_Z}  \left[{\overline v}(p_2) \gamma_\mu (4 \sin^2(\theta_W) - 1 + \gamma^5)  u(p_1)\right]  \times (k \Theta)^\mu \left[1 + \frac{i}{2} (p_2 \Theta p_1)\right]. 
\eea
Here $s=k^2$ with $k = p_1 + p_2 = p_3 + p_4$ and $s_Z= s - m_Z^2 + i \Gamma_Z m_Z$. $\alpha = e^2/4\pi$ and $\theta_W$ is the Weinberg angle. $m_h $ is the Higgs mass, $m_Z$ and $\Gamma_Z$ are the mass and decay width of the $Z$ boson. 
Using Eq.~\ref{unpol_photon} and Eq.~\ref{unpol_Z}, we find the spin averaged squared-amplitude as  
\bea \label{AmpSqrd}
\overline {|{\mathcal{A}}|^2} = \overline {|{\mathcal{A}_\gamma}|^2} + \overline {|{\mathcal{A}_Z}|^2} +
\overline {2 Re({\mathcal{A}^\dagger_\gamma} {\mathcal{A}_Z})} = \frac{1}{4} \sum_{spin} |{\mathcal{A}}|^2,
\eea
where several terms of Eq.~(\ref{AmpSqrd}) are given in Appendix C. 

 The noncommutative parameter $\Theta_{\mu\nu}$ is considered to be a fundamental constant in nature. It's direction is fixed with respect to an inertial(non rotating) coordinate system (which can be a celestial coordinate system). Now our experiment is done in the laboratory coordinate system which is located on the surface of the earth and is moving by the earth's rotation. This gives rise to an apparent time variation of the components of  $\Theta_{\mu\nu}$ which should be taken into account while making any phenomenological investigation.  

The effect of earth's rotation on noncommutative phenomenology were considered in several earlier studies \cite{NCearthrot}. Here we will follow Kamoshita's work \cite{NCearthrot} for the notations.
\begin{figure}[htbp]
\centerline{\hspace{-12.3mm}
{\epsfxsize=11cm\epsfbox{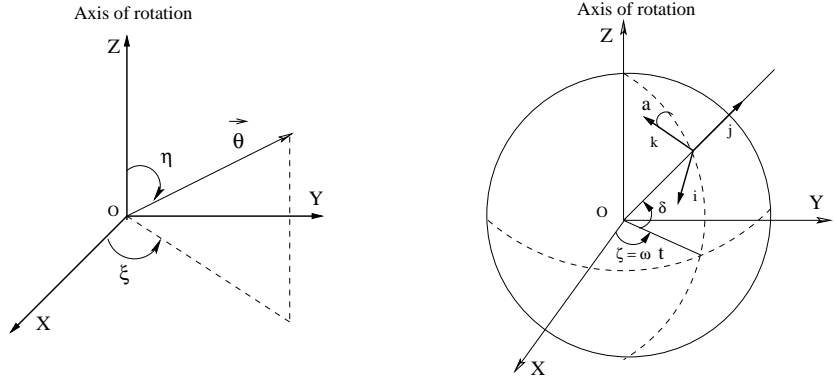}}}
\vspace*{-0.25in}
\caption{{\it In the left figure the primary coordinate system(X-Y-Z) is shown. The generic NC vector ${\vec{\Theta}}$ (electric or magnetic type)  of $\Theta_{\mu\nu}$ is shown with $\eta$ and 
$\xi$, respectively the polar and the azimuthal angle. In the right figure the arrangement of laboratory coordinate system (i-j-k) for an experiment on the earth in the primary coordinate system (X-Y-Z) is shown. In the above $\zeta = \omega t$ where $\omega$ is a constant. Also ($\delta$,~$a$), which defines the location of the laboratory, are constants.}}
\protect\label{sigplot}
\end{figure}
 Let ${\hat{i}}_X,~{\hat{j}}_Y$ and ${\hat{k}}_Z$ be the orthonormal bases of the primary(non rotating) coordinate system (X-Y-Z). Then in the laboratory coordinate system ($\hat{i} - \hat{j} - \hat{k}$), the bases vectors of the primary(non rotating) coordinate system can be written as  
\begin{eqnarray}
{\hat{i}}_X = \left(\begin{array}{c}
c_a s_\zeta + s_\delta s_a c_\zeta \\ c_\delta c_\zeta \\ s_a s_\zeta - s_\delta c_a c_\zeta
\end{array} \right), \nonumber 
{\hat{j}}_Y = \left(\begin{array}{c}
-c_a c_\zeta + s_\delta s_a s_\zeta \\ c_\delta s_\zeta \\ -s_a c_\zeta - s_\delta c_a s_\zeta
\end{array} \right), \nonumber 
{\hat{k}}_Z = \left(\begin{array}{c}
-c_\delta s_a  \\ s_\delta \\ c_\delta c_a 
\end{array} \right). \nonumber 
\end{eqnarray}
Here we have used the abbreviations $c_\delta = cos\delta,~s_\delta = sin\delta$ etc. In Fig. \ref{sigplot}, the primary($X - Y - Z$) and the laboratory($i - j - k$) coordinate system are shown. 
Note that the primary $Z$ axis is along the axis of earth's rotation and ($\delta, a$) defines the location of $e^- - e^+$ experiment on the earth, with $- \pi/2 \le \delta \le \pi/2$ and $0 \le a \le 2 \pi$.  Because of earth's rotation the angle $\zeta$ (see Fig. 2) increases with time and the  detector comes to its original position after a cycle of one complete day, one can define $\zeta = \omega t$ with $\omega = 2 \pi/T_{day}$ and $T_{day} = 23h56m4.09053s$.

 Using all these, the electric and the magnetic components of the NC parameter $\Theta_{\mu\nu}$ in the primary system is given by 
\bea
{\vec{\Theta}}_E = \Theta_E \sin\eta_E ~\cos\xi_E ~{\hat{i}}_X + \Theta_E \sin\eta_E ~\sin\xi_E ~{\hat{j}}_Y + \Theta_E  \cos\eta_E ~{\hat{k}}_Z \\
{\vec{\Theta}}_B = \Theta_B \sin\eta_B ~\cos\xi_B ~{\hat{i}}_X + \Theta_B \sin\eta_B ~\sin\xi_B ~{\hat{j}}_Y + \Theta_B \cos\eta_B ~{\hat{k}}_Z
\eea 
with
\beq
{\vec{\Theta}}_E = (\Theta^{01}, \Theta^{02}, \Theta^{03}), ~~ {\vec{\Theta}}_B = (\Theta^{23}, \Theta^{31}, \Theta^{12})
\eeq
and
\beq
\Theta_E = |{\vec{\Theta}}_E| = 1/\Lambda^2_E, ~~\Theta_B = |{\vec{\Theta}}_B| = 1/\Lambda^2_B.
\eeq
Here $(\eta, \xi)$ specifies the direction of the NC parameter $\Theta_{\mu\nu}$ w.r.t the primary coordinate system with $0 \le \eta \le \pi$ and $0 \le \xi \le 2 \pi$. In above $\Theta_E$ and $\Theta_B$ are the model parameters and the energy scales defined by $\Lambda_E = 1/\sqrt{\Theta_E}$ and $\Lambda_B = 1/\sqrt{\Theta_B}$  can be probed for different processes.

 The spin-averaged squared-amplitude of the $ e^+ e^- \stackrel{\gamma,Z}{\longrightarrow} \mu^+ \mu^-$ scattering is given by
\beq \label{Ampsq}
{\overline {|{\mathcal{A}}|^2}} = {\overline {|{\mathcal{A}}_\gamma|^2}} + {\overline {|{\mathcal{A}}_Z|^2}} + 2 {\overline {Re({\mathcal{A}}_Z {\mathcal{A}}_\gamma^{ \dagger })}}. 
\eeq
The direct and interference terms in Eq.~\ref{Ampsq} are given in Appendix C.  Since it is difficult to get the time dependent data, we take the average of the cross section or it's distribution over the sidereal day $T_{day}$. We introduce the time averaged observables as follows: 
\bea \label{dsigma_avg}
\left<\frac{d^2\sigma}{d\cos\theta~d\phi}\right>_T &=& \frac{1}{T_{day}} \int_{0}^{T_{day}} \frac{d\sigma}{d\cos\theta~d\phi} dt, \\
\left<\frac{d\sigma}{d\cos\theta}\right>_T &=& \frac{1}{T_{day}} \int_{0}^{T_{day}} \frac{d\sigma}{d\cos\theta} dt, \\
\left<\frac{d\sigma}{d\phi}\right>_T &=& \frac{1}{T_{day}} \int_{0}^{T_{day}} \frac{d\sigma}{d\phi} dt, \\
\left<\sigma\right>_T &=& \frac{1}{T_{day}} \int_{0}^{T_{day}} \sigma dt,
\eea
where
\bea 
\label{sigma}
\sigma &=& \int_{-1}^1 d(\cos\theta) \int_0^{2 \pi} d\phi \frac{d \sigma}{d\cos\theta~d\phi}, \\
\label{dsdcostheta}
\frac{d\sigma}{d\cos\theta} &=& \int^{2 \pi}_0 d\phi \frac{d \sigma}{d\cos\theta~d\phi},  \\
\label{dsdphi}
\frac{d\sigma}{d\phi} &=& \int^1_{-1} d(\cos\theta) \frac{d \sigma}{d\cos\theta~d\phi}. 
\eea
In the above
\beq 
\frac{d^2 \sigma}{d\cos\theta~d\phi} = \frac{1}{64 \pi^2 s} \left(1 - \frac{4 m_h^2}{s}\right)^{1/2}{\overline {|{\mathcal{A}}|^2}}, 
\eeq
where $\sigma$ = $\sigma(\sqrt{s}, \Lambda, \theta, \phi, t)$. The time dependence in the cross section or it's distribution enters through the NC parameter ${\vec{\Theta}}(={\vec{\Theta}}_E)$ which changes with the change in $\zeta = \omega t $. The angle parameter $\xi$ appears in the expression of $\vec{\Theta}$ through $\cos(\omega t - \xi)$ or 
$\sin(\omega t - \xi)$ \cite{NCearthrot} as the initial phase for time evolution gets disappeared in the time averaged observables. So one can deduce ${\vec {\Theta}}_E $ i.e. $\Lambda_E$ and the angle $\eta_E$ from the time-averaged observables.

\section{Numerical Analysis}
 Before making a detailed analysis, let us make some general remarks regarding the observation of noncommutative effects. Since we assume $c_{\mu\nu}=(c_{0i},c_{ij})=(\xi_i,~\epsilon_{ijk}\chi^k)$, where $\xi_i = (\vec{E})_i$ and $\chi_k = (\vec{B})_k$ are constant vectors in a frame that is stationary with respect to fixed stars, the vectors $(\vec{E})_i$ and $(\vec{B})_k$ point in fixed directions which are the same in all frames of reference. However, as the earth rotates around its axis and revolves around the Sun, the direction of $\vec{E}$ and $\vec{B}$ will change continuously with time dependence which is a function of the coordinates of the laboratory. The observables that are measured will thus show a characteristic time dependence. It is important to be able to measure this time dependence to verify such non commutative  theories. In one of our earlier work on pair production of Higgs boson, we have assumed the vectors $\vec{E}= \frac{1}{\sqrt{3}} (\hat{i} + \hat{j} + \hat{k}) $ and $\vec{B}= \frac{1}{\sqrt{3}} (\hat{i} + \hat{j} + \hat{k})$ i.e. they behave like constant vectors \cite{pkdas_higgs}, which happens to be the case at some instant of time. In the present work, we analyze in detail the effect of earth's rotation. We probe the characteristic NC scale $\Lambda$ and the orientation angle $\eta(=\eta_E)$  using the  time-averaged observables defined in the laboratory coordinate system. We set the laboratory coordinate system by taking $(\delta,~a)=(\pi/4,~\pi/4)$ which is the OPAL experiment at LEP. 

\subsection{Pair production cross section in the NCSM}
As we have mentioned earlier that the pair production of Higgs boson is forbidden in the SM, any signature of this event will correspond to new physics and the non-minimal NCSM is one of the promising candidates among these class of new physics models.
\begin{figure}[htbp]
\vspace{5pt}
\centerline{\hspace{-3.3mm}
{\epsfxsize=6cm\epsfbox{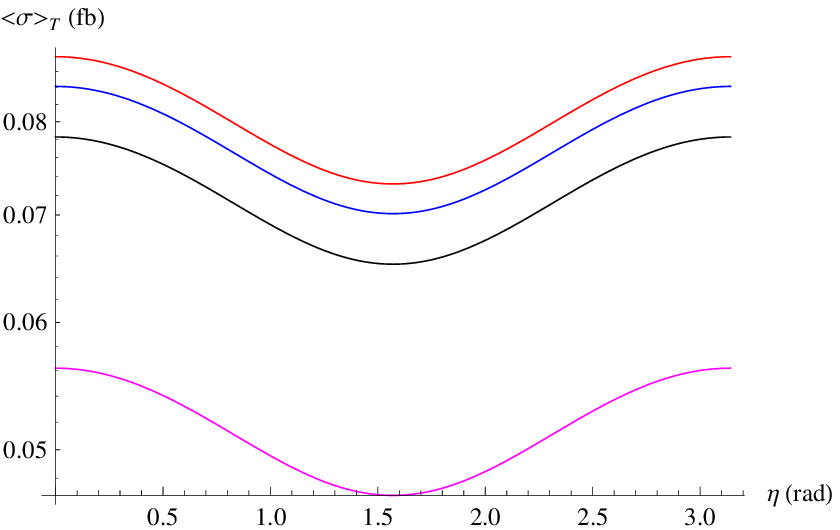}} \hspace{0.15in} {\epsfxsize=6cm\epsfbox{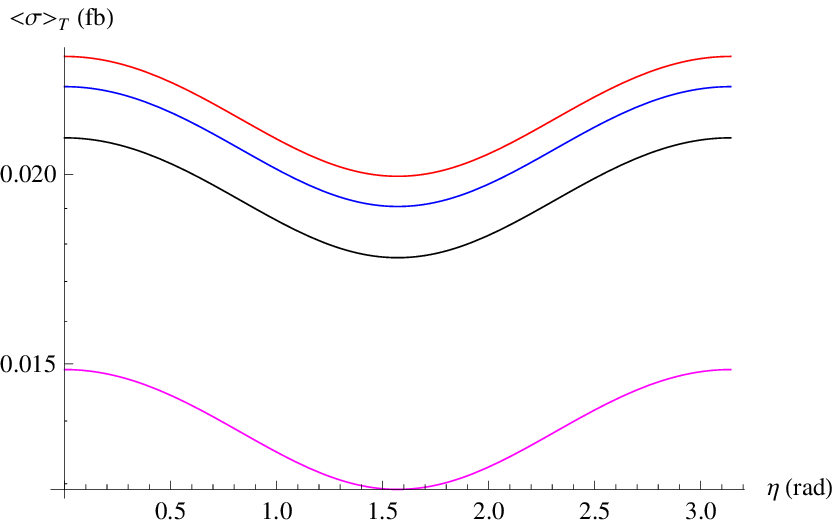}} }
\caption{\it{The time-averaged cross section $\left<\sigma(e^- e^+ \stackrel{\gamma,Z}{\rightarrow}\to H~H)\right>_T$ (fb) as a function of the orientation angle $\eta(~\eta_E)$ of the NC vector ${\vec{\Theta}}_E$ is shown.  On the left(right) panel the machine energy($E_{com}$) is fixed at $E_{com}=500(1000) ~\rm{GeV}$. While going from top to bottom, $m_h$ changes as $129,~127.5,~125 ~\rm{and} ~114.3~\rm{GeV}$, respectively. }}
\protect\label{sigma_eta}
\end{figure}
In Fig. \ref{sigma_eta}, we have plotted $\left<\sigma\right>_T$ (here $\left<\sigma(e^- e^+ \stackrel{\gamma,Z}{\rightarrow}\to H~H)\right>_T = \left<\sigma\right>_T $) as a function of the orientation angle 
$\eta$ of the NC vector ${\vec{\Theta}}_E$ corresponding to different Higgs mass at different machine energies. On the left(right) panel, the topmost, next to the topmost, next to it and the lowermost curves correspond to $m_h = 129,~127.5,~125~ \rm{and} ~114.3 ~\rm{GeV}$, respectively where the machine energy is being set at $E_{com}(=\sqrt{s})= 500~(1000)$ GeV. We see that  $\left<\sigma\right>_T$ is larger for higher  Higgs mass ($m_h$) and each plot (corresponding to different $m_h$ value) has a common minimum at $\eta = \pi/2$. The cross-section $\left<\sigma\right>_T$ is maximum at $\eta = 0$ and $\pi$, respectively. All these plots show the strong dependence of the pair production cross-section on the orientation of the NC vector i.e. on $\eta$ which will be probed in detail in the subsequent plots. 

\begin{figure}[htbp]
\centerline{\hspace{-12.3mm}
{\epsfxsize=7cm\epsfbox{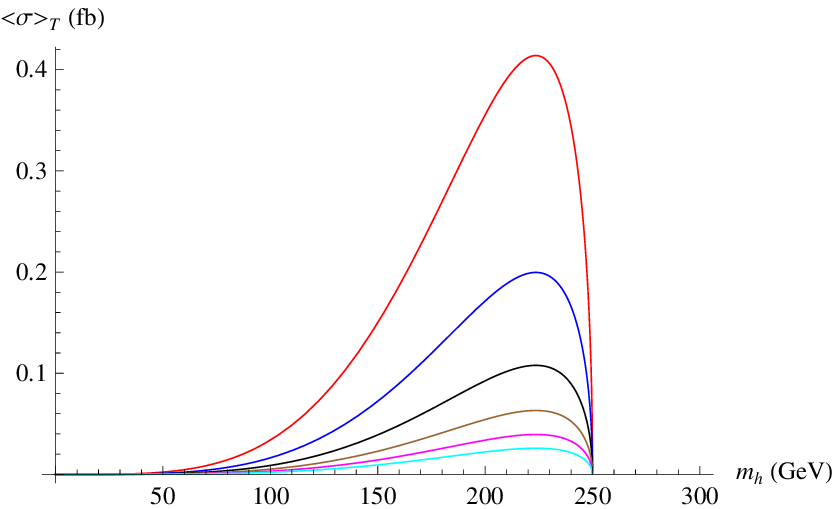}} \hspace{0.15in} {\epsfxsize=7cm\epsfbox{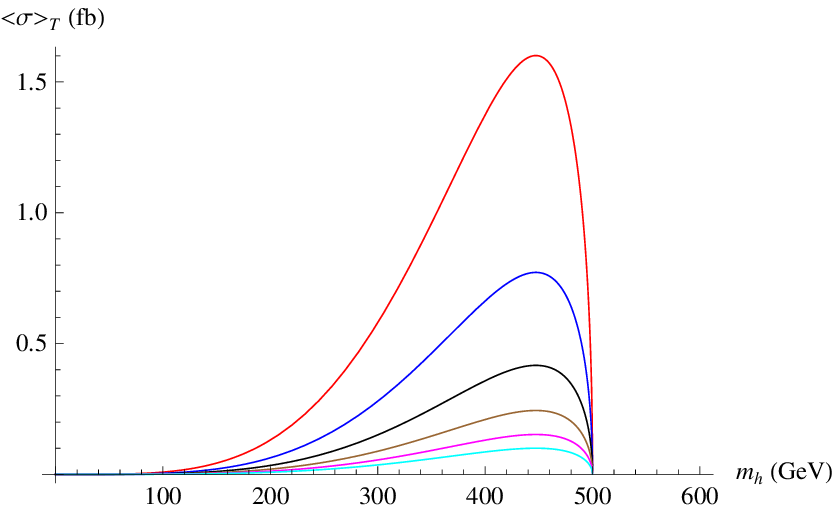}} }
\caption{\it{The time-averaged cross section $\left<\sigma \right>_T$ (fb) is shown as a function of the Higgs mass $m_h$. On the left(right) panel, the machine energy is fixed at $E_{com} = 500(1000)$~GeV. While going from the top to bottom the NC scale $\Lambda$ increases from $500$ GeV to $1000$ GeV in steps of $100$ GeV.}}
\protect\label{sigma_higgsmass}
\end{figure}

In Fig. \ref{sigma_higgsmass}, the time-averaged total cross section $\left<\sigma\right>_T$ is plotted as a function of Higgs mass $m_h$. In the left(right) panel, the machine energy is fixed at $E{com}= 500~(1000) ~\rm{GeV}$, respectively.  
In each panel while going from the top to the bottom, the curves correspond to $\Lambda = 500,~600,~700,~800,~900$ and $1000$~GeV, respectively and for all curves $\eta$ is fixed at $0(\pi)$. Note that the pair production cross section is maximum at $m_h = 224(447)$ GeV corresponding to the machine energy $E_{com}= 500(1000)$ GeV.

\vspace*{-0.2in} 
\begin{center}
Table 1
\end{center}
\vspace*{-0.15in}
\begin{center}
\begin{tabular}{|c|c|c|c|c|c|c|c|c|c|c|}
\hline
$E_{com}$ & $\Lambda$ & $\sigma$(fb) & ${\mathcal{L}}(fb^{-1})$  & N ($yr^{-1}$)&& $E_{com}$ & $\Lambda$ & $\sigma$(fb) & ${\mathcal{L}}(fb^{-1})$  & N ($yr^{-1})$\\
\hline
   & 500 &0.4139 & 500 & 207 &&&500 &1.6008 & 500& 800\\
  & 600 & 0.1996 & 500 & 100 &&&600&0.7720 &500& 386\\
  500 & 700 & 0.1077 & 500 &54 &&1000&700& 0.4167&500& 208\\
  & 800 & 0.0632& 500 & 31 &&&800& 0.2443&500& 122\\
  & 900 & 0.0394 & 500 & 20 &&&900& 0.1525&500& 76\\
  & 1000 & 0.0259 & 500 & 13 &&&1000& 0.1000&500& 50\\
\hline
\end{tabular}
\end{center}
\noindent {\it Table 1: The yearly number of events(yielding NC signals) with the increase in $\Lambda$ (in GeV) are shown. The integrated luminosity of the LC is assumed to be ${\mathcal{L}}=500~fb^{-1}$. The orientation angle $\eta$ of the NC vector is chosen to be $0(\pi)$. The machine energy $E_{com} (=\sqrt{s})$ is expressed in GeV. }

 Assuming an integrated luminsoity $\mathcal{L} = 500~fb^{-1}$ (per year) of the futute Linear Collider, we predict the number of events of Higgs pair production in the NCSM which are shown in Table 1 as function of $\Lambda$ corresponding to the machine energy $\sqrt{s}=500$ GeV and $1000$ GeV, respectively(see above). In Table 1, we see that as $\Lambda$ increases from $500$ GeV to $1000$ GeV, the number of pair production events $N (yr^{-1})$ decreases from $207(800)$ per year to  $13(50)$ per year for the machine energy $E_{com}=500~(1000) ~\rm{GeV}$. So a maximum of $207$ and $800$ events (NC signal) per year are expected to be observed at the upcoming linear collider correponding to the machine energy $E_{com} = 500$ GeV and $1000$ GeV with the LC luminosity ${\mathcal{L}}=500~fb^{-1}$.

\begin{figure}[htbp]
\vspace{0.25in}
\centerline{\hspace{-12.3mm}
{\epsfxsize=7cm\epsfbox{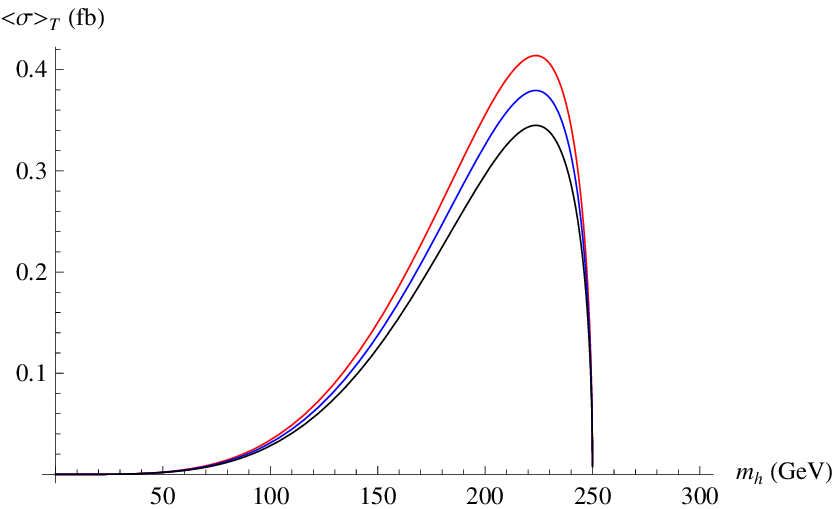}} \hspace{0.15in} {\epsfxsize=7cm\epsfbox{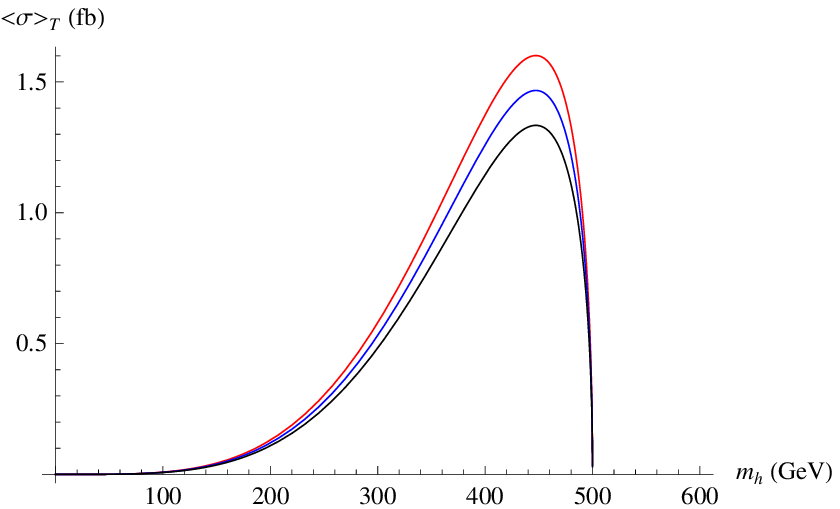}} }
\vspace*{-0.2in}
\caption{\it{The time-averaged cross section $\left<\sigma\right>_T$ (fb) is plotted as a function of $m_h$. On the left(right) panel, the machine energy is fixed at $E_{com} =500(1000)$~GeV. While going from the top to the bottom curve the orientation angle $\eta$ changes from $0(\pi)$  to $\pi/2$ in steps of $\pi/4$. The NC scale $\Lambda$ is being set at $500$ GeV.}}
\protect\label{sigma_higgs_eta}
\end{figure}

 In Figure \ref{sigma_higgs_eta} we have plotted $\left<\sigma\right>_T$ as a function of Higgs mass $m_h$ corresponding to different $\eta$. The left and right panels respectively correspond to $E_{com}= 500$ GeV and $1000$ GeV. In each panel the topmost curve corresponds to $\eta = 0$ or $\pi$, the middle one corresponds to $\eta = \pi/4$ or $3 \pi/4$ and the lowermost curve corresponds to  $\eta = \pi/2$.  In the left(right) panel, the cross-section decreases from $0.414(1.601)$ fb to $~0.345(1.334)$ fb as $\eta$ changes from $0$ to $\pi/2$ at $m_h=224$ GeV and $447$ GeV corresponding to $\Lambda = 500$ GeV.
\vspace{-0.3in}
\subsection{Total cross-section as a function of the machine energy in the NCSM}
\begin{figure}[htbp]
\vspace{-0.1in}
\centerline{\hspace{-12.3mm}
{\epsfxsize=7cm\epsfbox{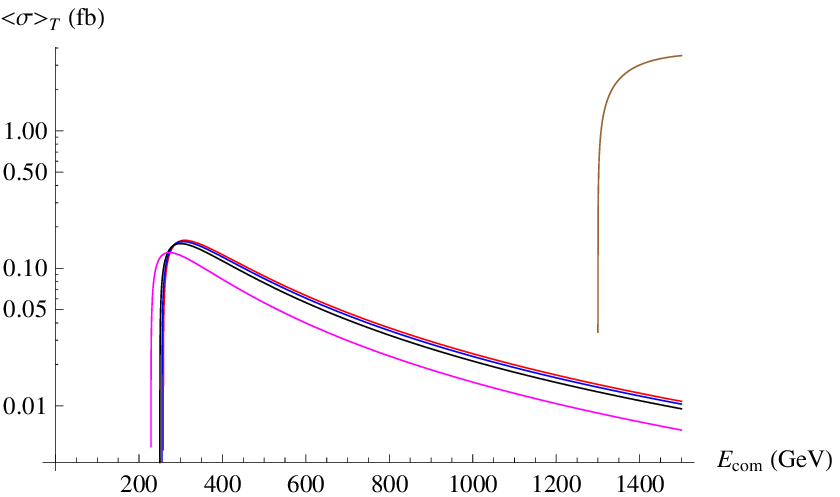}} \hspace{0.15in} {\epsfxsize=7cm\epsfbox{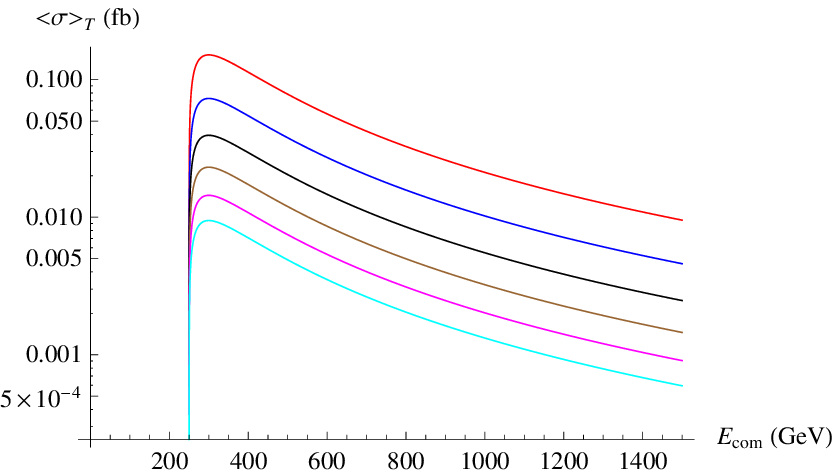}}}
\vspace*{-0.2in}
\caption{\it{The time-averaged cross-section $\left<\sigma \right>_T$ (fb) is plotted against the machine energy $E_{com}$. On the left panel while going from the topmost to the lowermost curve(among four closely packed curves) $m_h$ decreases as $129,~127.5,~125 ~\rm{and}~114.3$ GeV, respectively. The single isolated curve on the extreme right corresponds to a heavy Higgs of mass $m_h = 650 ~\rm{GeV}$ (allowed by both ATLAS and CMS data). In this plot we choose $\Lambda = 500$ GeV and $\eta=0(\pi)$. On the right panel, 
$\Lambda$ increases from $500$ GeV to $1000$ in steps of $100$ GeV as one moves from the topmost to the lowermost curve. In this plot we set $\eta = 0$ and $m_h = 125~\rm{GeV}$for this plot.}}
\protect\label{sigma_energy}
\end{figure}

\vspace{-0.1in}
In Figure \ref{sigma_energy} we have shown the time-averaged total cross-section $\left<\sigma\right>_T$ as a function of the machine energy $E_{com}$. 
On the left panel the plots for different $m_h$ with $\Lambda = 500$ GeV and $\eta = 0 (\pi)$ are shown. On the right panel, the plots corresponding to Higgs mass $m_h =125$ GeV and for different $\Lambda$ ranging from $500$ GeV to $1$ TeV are shown. Among the four closely packed curves(on the left panel), the topmost, next to the topmost (below), next to that and the lowermost curves respectively correspond to $129,~127.5,~125 ~\rm{and}~114.3$ GeV.  The isolated single curve(on the top right)  corresponds to $m_h = 650~\rm{GeV}$ (which is a value allowed by ATLAS and CMS data). The cross-section $\left<\sigma \right>_T$ of the pair production of $125$ GeV mass Higgs boson changes from $0.0783$ fb to $0.0211$ fb (the number of events $N (yr^{-1})$ changes from $39$(per year) to $11$(per year)) as the machine energy changes from $500$ GeV to $1000$ GeV with the machine luminosity  ${\mathcal{L}}=500~fb^{-1}$ (per year).
On the right panel, the topmost(lowermost) curve correponds to $\Lambda = 500(1000)$ GeV. From the top to the bottom $\Lambda$ increases from $500$ GeV to $1000$ GeV in steps of $100$ GeV. The cross-section $\left<\sigma \right>_T$ varies from $0.0783$ fb to $0.0049$ fb (so the number of events changes from $39$(per year) to $2$(per year)) at the machine energy  $E_{com}=500$ GeV with ${\mathcal{L}}=500~fb^{-1}$ (per year).

\begin{figure}[htbp]
\vspace{-0.10in}
\centerline{\hspace{-12.3mm}
{\epsfxsize=7cm\epsfbox{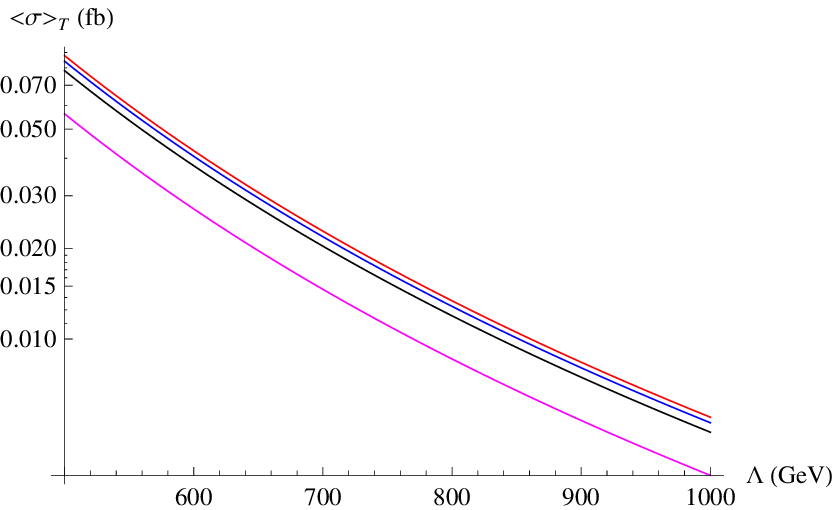}} \hspace{0.15in} {\epsfxsize=7cm\epsfbox{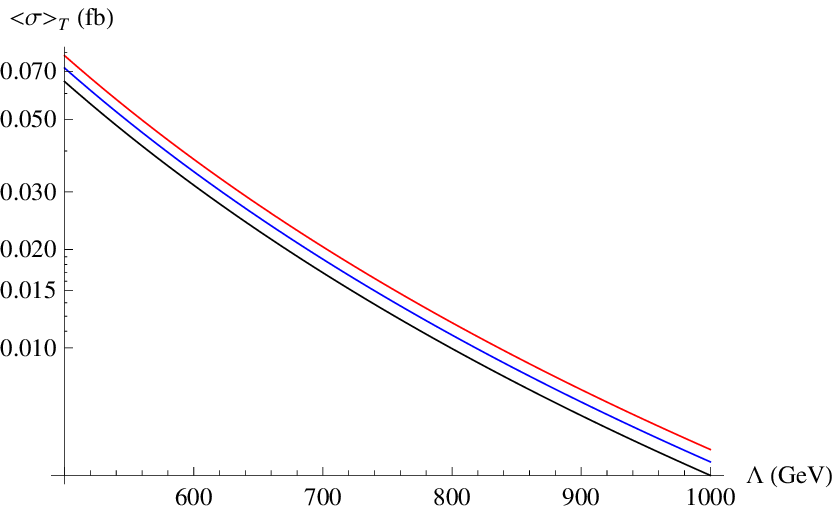}} }
\vspace*{-0.1in}
\caption{{\it{The cross-section $\left<\sigma \right>_T$ is plotted as a function of the NC scale $\Lambda$ corresponding to different $m_h$(left panel) with $\eta = 0$.  For the topmost, next to the topmost, next to it and the lowermost curve, $m_h$ corresponds to $129,~127.5,~125 ~\rm{and}~114.3$ GeV. On the right panel we plot the cross section against $\Lambda$ for different $\eta$ but with $m_h = 125~\rm{GeV}$}}.}
\protect\label{sigma_NCscale}
\end{figure}

 It may be worthwhile to see how the cross-section depends on the NC scale $\Lambda$ corresponding to different orientation angles $\eta$ and Higgs mass $m_h$. In Fig.~\ref{sigma_NCscale} we have shown this. On the left panel of this Figure the cross-section $\left<\sigma \right>_T$ is plotted against $\Lambda$ for $\eta = 0$ and $m_h = 129,~127.5,~ 125 ~\rm{and}~114.3 \rm{GeV}$ for the topmost, next to the topmost, next to that and the lowermost curves, respectively. On the right panel, $\left<\sigma \right>_T$ is plotted against $\Lambda$ for $m_h = 125$ GeV and $\eta = 0~(\rm{for~the ~topmost ~curve}), \pi/4~(\rm{for ~the ~middle ~curve})$ and $\pi/2~(\rm{for ~the ~lowermost ~curve})$. For both plots, we set the machine energy at $E_{com} = 500~\rm{GeV}$. 
\subsection{Time varying total cross-section  }
\begin{figure}[htbp]
\vspace{0.25in}
\centerline{\hspace{-12.3mm}
{\epsfxsize=7cm\epsfbox{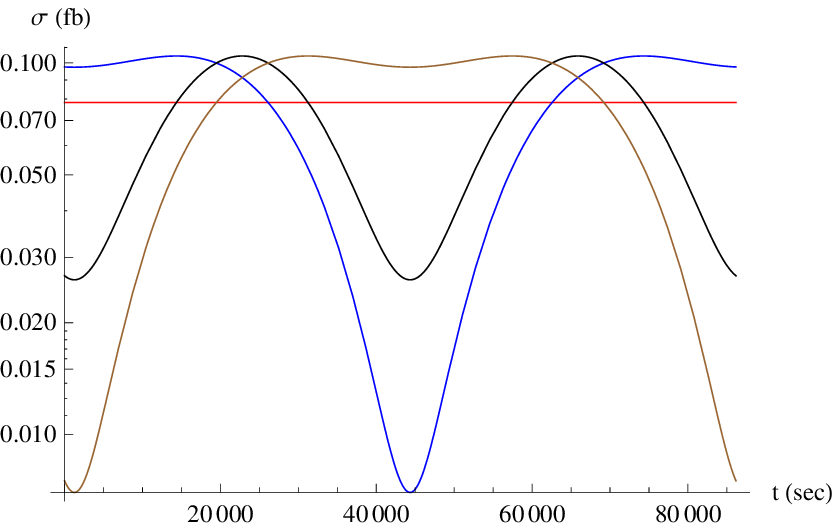}} \hspace{0.15in} {\epsfxsize=7cm\epsfbox{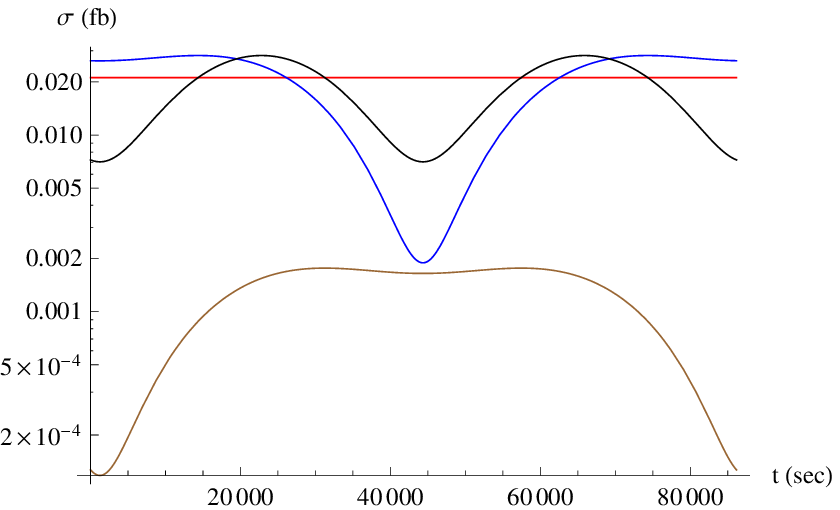}} }
\caption{The variation of the time-dependent cross section $\sigma(e^- e^+ \stackrel{\gamma,Z}{\rightarrow}\to )$ (fb) with time(t) over a complete day ($T_{day}$) is shown. For the left(right) panel the machine energy is being set at $E_{com} = 500(1000)$ GeV and the Higgs mass at $m_h = 125$ GeV. We have chosen 
$\Lambda$ to be $500$ GeV and $\eta$ as $0,~\pi/4,~\pi/2$ and $3 \pi/4$ (see in the text below)}
\protect\label{sigma_time}
\end{figure}

Finally, we are to see how the production cross-section varies over a complete day ($T_{day}$). In Fig.~\ref{sigma_time} we have shown how the cross-section varies with time over a complete day correponding to $\eta = 0$ (horizontal red line), $\pi/4$ (blue curve), $\pi/2$ (black curve) and $3 \pi/4$ (brown curve) and $\Lambda = 500$ GeV. For left(right) panel we have set the machine energy at $E_{com} = 500(1000)$ GeV and the Higgs mass at $m_h = 125$ GeV.  
Interestingly, the pair production cross-section have several maxima and minima during different times of the day. The pattern of each plot strongly depends on the value of $\eta$, the orientation angle of the NC (electric) vector with the axis of earth's rotation and the value of $m_h$ (which is chosen to be $125$ GeV for this Figure).
\vspace*{-0.1in}
\subsection{Constraining the NC scale $\Lambda$ using experimental bound on Higgs mass $m_h$}
We found in an earlier subsection that the pair production cross section is maximum at $m_h = 224$ GeV and $447$ GeV corresponding to the c.o.m energy $E_{com} = 500$ GeV and $1000$ GeV with the NC scale $\Lambda = 500$. 
\begin{figure}[htbp]
\centerline{\hspace{-12.3mm}
{\epsfxsize=7cm\epsfbox{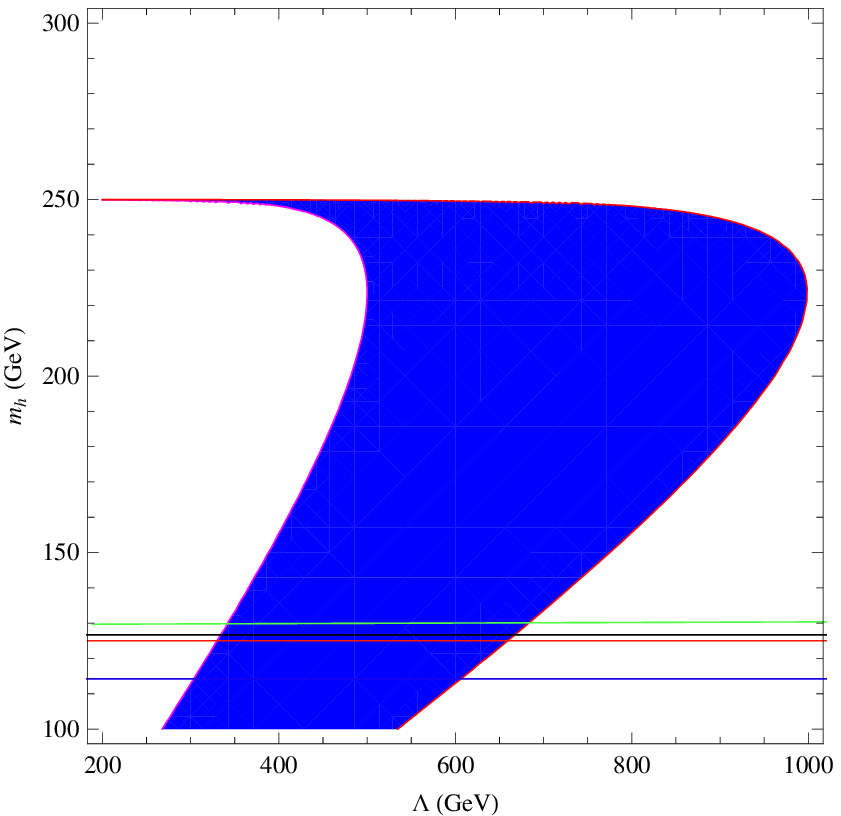}} \hspace{0.15in} {\epsfxsize=7cm\epsfbox{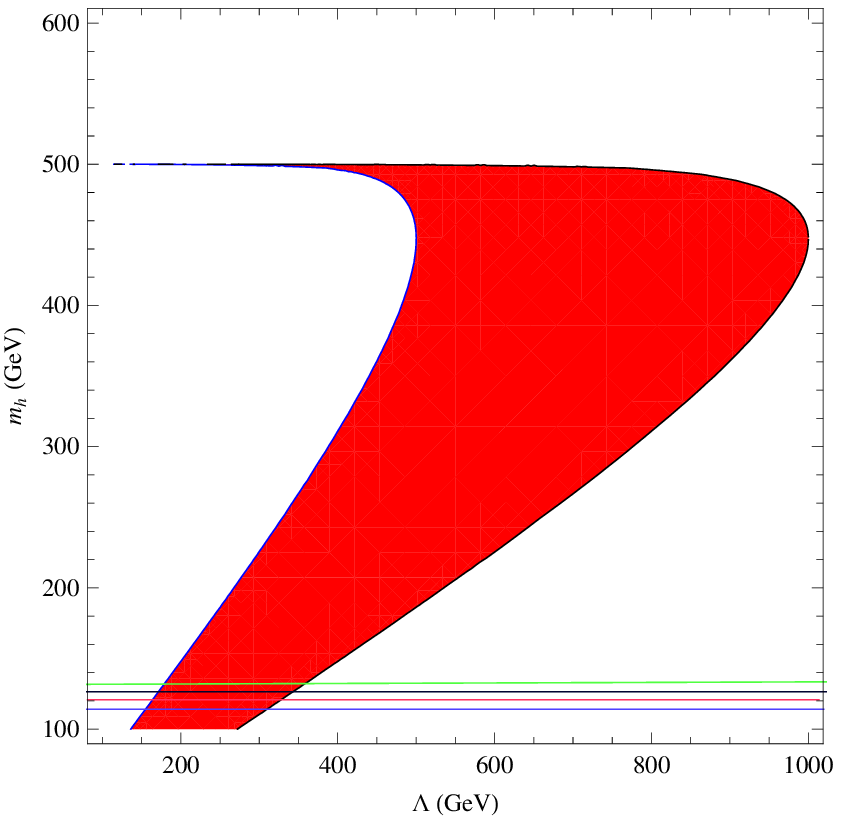}} }
\vspace*{-0.15in}
\caption{\it{The contour plot in the $m_h - \Lambda$ plane obtained by setting $N_{min} \le N(yr^{-1}) \ge N_{max}$ at a certain machine energy $E_{com}$. On the left panel, $E_{com} = 500$ GeV and $\eta = 0(\pi)$.  The region shaded in blue colour and bounded by the contour curves corresponding to $N(=N_{max}) = 13$ (on the right) and  $N(=N_{max}) = 207$ (on the left) is allowed. On the right panel, $E_{com} = 1000$ GeV and $\eta = 0(\pi)$.  The region shaded in red colour and bounded by the contour curves corresponding to $N(=N_{max}) = 50$ (on the right) and  $N(=N_{max}) = 800$ (on the left) is allowed. In each panel the  lowermost horizontal line correspond to the lower bound $m_h(\sim 114.3)$ GeV which follows from the LEP II direct search of Higgs boson. While going from the bottom to the top, the next to the lowermost horizontal line correspond to $m_h = 125$ GeV (the most promising value of Higgs mass as reported by LHC). The topmost curve and next to it correspond to $m_h = 129 ~\rm{GeV}$ and $127.5~\rm{GeV}$, the lower limit of the excluded Higgs mass as reported by ATLAS and CMS collaborations, respectively.}}
\protect\label{contourplot}
\end{figure}
In Figure \ref{contourplot}  we make a contour plot in the $m_h - \Lambda$ plane corresponding to following event production rate per year $N (yr^{-1})$: 
\begin{itemize}
\item Scenario I : $13 \le N~(\rm{yr^{-1}})  \le 207$ with the machine energy $E_{com} = 500$ GeV,
\item Scenario II: $50 \le N~(\rm{yr^{-1}})   \le 800$ with the machine energy $E_{com} = 1000$ GeV.
\end{itemize} 
The following results are in the order:
\begin{itemize}
\item The direct search of Higgs boson at LEP II gives a lower bound on Higgs mass $m_h$ as $114.3$ GeV. Incorporating this in 
Fig. \ref{contourplot}, one finds : (i)$ 303 ~\rm{GeV} \le \Lambda \le 606 ~\rm{GeV}$ in scenario I and (ii) $ 155 ~\rm{GeV} \le \Lambda \le 328 ~\rm{GeV}$ in scenario II. 
\item Both ATLAS and CMS collaborations of LHC find tantalizing hints for a Higgs boson of mass $m_h = 125 \rm{GeV}$ and using $125$ Higgs mass value we obtain the following constraints on $\Lambda$: (i) $ 330 ~\rm{GeV} \le \Lambda \le 660 ~\rm{GeV}$ in scenario I and (ii) $ 169 ~\rm{GeV} \le \Lambda \le 339 ~\rm{GeV}$ in scenario II.
\item ATLAS and CMS collaborations excludes the SM Higgs boson of mass $129 ~\rm{GeV} \le m_h \le 539 ~\rm{GeV}$ (ATLAS) and $127.5 ~\rm{GeV} \le m_h \le 600 ~\rm{GeV}$ (CMS). For 
$m_h = 129(127.5) ~\rm{GeV}$ we find the bound on $\Lambda$ as: (i)$ 339(336) ~\rm{GeV} \le \Lambda \le 677(670) ~\rm{GeV}$ in scenario I and (ii) $ 175(173) ~\rm{GeV} \le \Lambda \le 349(346) ~\rm{GeV}$ in scenario II. 
\end{itemize}
\section{Conclusion}
We have studied the pair production of Standard Model Higgs boson in the background of noncommutative space-time using the effect of earth rotation. Working within the non-minimal NCSM, we find that the production cross-section(time-averaged) strongly depends on the the orientation angle $\eta (=\eta_E)$ of the NC electric vector ${\vec{\Theta}}_E$ defined in the primary system: it is maximum at $\eta = 0,\pi$ and is minimum at $\eta = \pi/2$. We plot the time-averaged total cross section $\left<\sigma\right>_T$  as a function of Higgs mass $m_h$ (for a wide range of Higgs mass)  at a fixed machine energy $E_{com}$ for different NC scale $\Lambda$: the cross section is maximum at $m_h = 224 \rm{GeV}$  and $m_h = 447 \rm{GeV}$ and are about $0.414 \rm{fb}$ and $1.601 \rm{fb}$, respectively corresponding to the machine energy $E_{com}=500(1000) \rm{GeV}$ for $\Lambda = 500 \rm{GeV}$. The cross-section $\left<\sigma\right>_T$ and the maximum number of events(N) per year decreases with the increase in $\Lambda$. For example corresponding to $E_{com} = 500 \rm{GeV}$ and taking the integrated luminosity of LC about $\mathcal{L} = 500 \rm{fb^{-1}}$, we see $N(yr^{-1})$ drops from $207~\rm{yr^{-1}}$ to $13~\rm{yr^{-1}}$ as $\Lambda$ falls from $500 \rm{GeV}$ to $1000 \rm{GeV}$ which can be compared with no events in the case of SM.
We found no non-trivial polar and azimuthal distribution of the Higgs pair production cross-section at the LC energy as observed in the case of fermion pair production. 
 We plot the cross-section $\left< \sigma \right>_T$ as a function of the machine energy $E_{com}$  corresponding to $m_h = 129,~127.5,~125~ \rm{and}~114.3~\rm{GeV}$. For a Higgs boson of mass $m_h = 125~\rm{GeV}$, we find events of $39 ~\rm{yr^{-1}}$ and 
$11 ~\rm{yr^{-1}}$ corresponding to the machine energy  $E_{com} = 500~ \rm{GeV}$ and $1000~ \rm{GeV}$, respectively. We also show how  $\left< \sigma \right>_T$ changes with the NC scale $\Lambda$ for $m_h = 125 ~\rm{GeV}$ and $\eta = 0(\pi)$. 
We plot the time-dependent cross section ($\sigma$) as a function of time $t$ over a complete day $T_{day}$ corresponding to different $\eta$ at a particular machine energy and find that it changes quite significantly with times in a day dependig on the value of $\eta$ and $m_h$.    
 Finally, we make the contour plots in the plane of $m_h - \Lambda$ corresponding to the event rate $13 \le N(yr^{-1}) \le 207$ and $50 \le N(yr^{-1}) \le 800$. Using the lower bound $m_h \ge 114.3 ~\rm{GeV}$ (LEP II lower bound on $m_h$) and $m_h = 125 ~\rm{GeV}$ (the most promising value of Higgs mass as reported by ATLAS and CMS collaborations of LHC at CERN), we find the following bound on the NC scale $\Lambda$: Case-I (from LEP II data) (i) $303~ \rm{GeV} \le \Lambda \le 606 ~\rm{GeV}$ and  Case-II (from LHC data) $330~ \rm{GeV} \le \Lambda \le 660 ~\rm{GeV}$ corresponding to the Linear Collider energy $E_{com} = 500 ~\rm{GeV}$.  
\begin{acknowledgments}
The work of P.K.Das is supported by the BITS SEED Grant Project 2011 and the DAE BRNS Project (Project Ref. No. 2011/37P/08/BRNS). 
\end{acknowledgments}
\appendix
\section{Feynman rules to order ${\mathcal{O}}(\Theta)$ }
\noindent The Feynman rule for the $f(p_{in})- f(p_{out}) - \gamma(k)$ vertex is \cite{Melic:2005ep}
\bea
i e Q_f \gamma_\mu + \frac{1}{2}e Q_f \left[(p_{out} \Theta p_{in}) \gamma_\mu - (p_{out} \Theta)_\mu  (\pinsla -  m_f) - (\poutsla -  m_f) (\Theta p_{in})_\mu\right].
\eea
and for the $f(p_{in})- f(p_{out}) - Z(k)$ vertex is 
\bea
\frac{e}{sin2\theta_W} \left[i \gamma_\mu \Gamma_A^- \right] + \frac{e}{2 sin2\theta_W} \left[(p_{out} \Theta p_{in}) \gamma_\mu \Gamma_A^- 
 - (p_{out} \Theta)_\mu \Gamma_A^+ (\pinsla -  m_f) - (\poutsla -  m_f) \Gamma_A^- (\Theta p_{in})_\mu \right]. \nonumber \\ 
\eea
\noindent Here $\Gamma_A^\pm = (c_V^e \pm c_A^e \gamma_5)$ and $ p_{out} \Theta p_{in} = p_{out}^\mu  \Theta_{\mu \nu}  p_{in}^\nu = -p_{in} \Theta p_{out}$. At the vertex the momentum conservation reads as $p_{in} + k = p_{out}$. Similarly, the Feynman rule for the interaction vertex $H(p_3)-H(p_4)-Z(k)$ is
\beq
\frac{g~ m_h^2~ (k \Theta)_\mu}{4 \cos\theta_W}
\eeq
and for the vertex $H(p_3)-H(p_4)-\gamma(k)$ is
\beq
\frac{e~ m_h^2~ (k \Theta)_\mu}{4}
\eeq
In the above expressions, $(k\Theta)_\mu = k^{\nu} \Theta_{\nu \mu} $.

\section{Momentum prescriptions and dot products}
We work in the center of momentum(c.o.m) frame where the 4 momenta of the incoming and outgoing particles are given by
\bea
\label{prescstart} p_1 &=& \left(\frac{\sqrt{s}}{2}, 0, 0, \frac{\sqrt{s}}{2}\right)\\
p_2 &=& \left(\frac{\sqrt{s}}{2}, 0, 0, -\frac{\sqrt{s}}{2}\right)\\
p_3 &=& \left(\frac{\sqrt{s}}{2},k' \sin\theta \cos\phi, k' \sin\theta \sin\phi, k' \cos\theta \right)   \\
p_4 &=& \left(\frac{\sqrt{s}}{2},-k' \sin\theta \cos\phi,-k' \sin\theta \sin\phi, -k' \cos\theta \right), \\
k' &=& \frac{\sqrt{s}}{2} \sqrt{1- \frac{4 m_h^2}{s} } \nonumber
\eea
where  $\theta$ is the scattering angle made by the $3$-momentum vector $p_3$ of $H(p_3)$ with the +ve $\hat{z} (= \hat{k})$ axi\vspace*{-0.1in}s(the $3$-momentum direction of the incoming electron) and $\phi$ is the azimuthal angle. 

 The antisymmetric NC tensor $\Theta_{\mu \nu} = ({\vec{\Theta}}_E,~{\vec{\Theta}}_B)$ i.e. it has $3$  electric and $3$ magnetic components. The $s$-channel driven muon pair production in electron-positron collision is found to be sensitive only to the ${\vec{\Theta}}_E$ vector and hence one obtain constraints on $\Lambda_E$ (= $\Lambda$, say). 
In the laboratory frame (with $\eta = \eta_E$,~$\xi = \xi_E$ ), the electric NC vector ${\vec{\Theta}}_E $ can be written as 
\bea
{\vec{\Theta}}_E  &=& \Theta_E \sin\eta ~\cos\xi ~{\hat{i}}_X + \Theta_E \sin\eta ~\sin\xi ~{\hat{j}}_Y + \Theta_E \cos\eta ~{\hat{k}}_Z \nonumber \\
                 &=& \Theta^{lab}_{Ex} {\hat{i}} + \Theta^{lab}_{Ey} {\hat{j}} + \Theta^{lab}_{Ez} {\hat{k}}
\eea 
where
\bea
\Theta^{lab}_{Ex} &=& \Theta_E \left( s_\eta c_\xi (c_a s_\zeta + s_\delta s_a c_\zeta) + s_\eta s_\xi (-c_a c_\zeta + s_\delta s_a s_\zeta) - c_\eta c_\delta s_a)  \right) \nonumber \\
\Theta^{lab}_{Ey} &=& \Theta_E \left( s_\eta c_\xi c_\delta c_\zeta + s_\eta s_\xi c_\delta s_\zeta + c_\eta s_\delta \right) \nonumber \\
\Theta^{lab}_{Ez} &=&  \Theta_E \left( s_\eta c_\xi (s_a s_\zeta - s_\delta c_a c_\zeta) - s_\eta s_\xi (s_a c_\zeta + s_\delta c_a s_\zeta) + c_\eta c_\delta c_a)  \right)
\eea
with (see the main section)
\beq
\Theta_E = |{\vec{\Theta}}_E| = 1/\Lambda^2.
\eeq
In  above we have used abbreviations viz $s_\eta = \sin\eta,~c_\xi = \cos\xi$ etc.  As mentioned before, $(\eta, \xi)$ specifies the direction of ${\vec{\Theta}}_E$ w.r.t the primary coordinate system with $0 \le \eta \le \pi$ and $0 \le \xi \le 2 \pi$.  Using these we find
\bea
p_2 \Theta p_1 &=&  -\frac{s}{2} \Theta^{lab}_{Ez}, \\
(k\Theta)_0 &=&  0,\\
(k\Theta)_1 &=& -\sqrt{s}~ \Theta^{lab}_{Ex},\\
(k\Theta)_2 &=& -\sqrt{s}~  \Theta^{lab}_{Ey},\\
(k\Theta)_3 &=& -\sqrt{s}~  \Theta^{lab}_{Ez},
\eea
where $\Theta^{lab}_{Ex},~\Theta^{lab}_{Ey}$ and $\Theta^{lab}_{Ez}$ are defined above. Noting $k = p_1 + p_2=p_3 + p_4$, we find $ p_1.p_2 = s/2 = p_3.p_4$ and $(k\Theta).(k\Theta) = - s \left(\vec{\Theta}^{lab}_{E}.\vec{\Theta}^{lab}_{E}\right)$.
\vspace*{-0.1in}
\section{Spin-averaged squared amplitude}
\vspace*{-0.1in}
\noindent The various components of Eq.(\ref{AmpSqrd}) are found to be 
\bea
\overline {|{\mathcal{A}_\gamma}|^2} &=& - \frac{\pi^2 \alpha^2 m_h^4}{s^2} {\mathcal{F}}, \\
\overline {|{\mathcal{A}_Z}|^2} &=& -\frac{\pi^2 \alpha^2 m_h^4}{\sin^4(2\theta_W)} \frac{[1+(4 \sin^2\theta_W -1)^2]}{[(s-m_Z^2)^2 + \Gamma_Z^2 m_Z^2]} {\mathcal{F}}, \\
\overline {2 Re({\mathcal{A}^\dagger_\gamma} {\mathcal{A}_Z})} &=& \frac{2 \pi^2 \alpha^2 m_h^4}{\sin^2(2\theta_W)} \frac{(4 \sin^2\theta_W -1) (s - m_Z^2)}{s [(s-m_Z^2)^2 + \Gamma_Z^2 m_Z^2]} {\mathcal{F}}.
\eea
The overall factor ${\mathcal{F}}$ is given by
\bea
{\mathcal{F}} = [2 (p_1 \Theta p_2)^2 + (p_1.p_2) ((k\Theta).(k\Theta))].
\eea
The dot product terms appearing above are listed in Appendix B.
\newpage


\begin{thebibliography}{10}
\bibitem{Joe} "The results are preliminary but the 5 sigma signal at around 125 GeV we’re seeing is dramatic. This is indeed a new particle. We know it must be a boson and it’s the heaviest boson ever found," said CMS experiment spokesperson Joe Incandela. "The implications are very significant and it is precisely for this reason that we must be extremely diligent in all of our studies and cross-checks."
\end{thebibliography}
\end{document}